\documentclass[12pt,preprint]{aastex}

\slugcomment{}

\shorttitle{Two More SDSS AM CVn}
\shortauthors{Anderson et al.}

\received{}
\begin{document}


\title{Two More Candidate AM Canum Venaticorum (AM~CVn) Binaries 
from the Sloan Digital Sky Survey \altaffilmark{1}}


\author{
Scott F. Anderson\altaffilmark{2},
Andrew C. Becker\altaffilmark{2},
Daryl Haggard\altaffilmark{2},
Jose Luis Prieto\altaffilmark{3},
Gillian R. Knapp\altaffilmark{4},
Masao Sako\altaffilmark{5},
Kelly B. Halford\altaffilmark{4,6},
Saurabh Jha\altaffilmark{7},
Blake Martin\altaffilmark{4},
Jon Holtzman\altaffilmark{8},
Joshua A. Frieman\altaffilmark{9,10},
Peter M. Garnavich\altaffilmark{11},
Suzanne Hayward\altaffilmark{2},
\v Zeljko~Ivezi\'c\altaffilmark{2},
Anjum S. Mukadam\altaffilmark{2},
Branimir Sesar\altaffilmark{2},
Paula Szkody\altaffilmark{2},
Viktor Malanushenko\altaffilmark{12},
Michael W. Richmond\altaffilmark{13},
Donald P. Schneider\altaffilmark{14},
Donald G. York\altaffilmark{15,16}
}



\altaffiltext{1}{Includes optical observations obtained with: 
the Sloan Digital Sky Survey I and II and
the Apache Point Observatory (APO) 3.5m telescope which is owned and 
operated by the Astrophysical Research Consortium (ARC);
the WIYN Observatory which is a joint facility of the Univ. of Wisconsin, Indiana 
Univ., Yale Univ., and NOAO.}

\altaffiltext{2}{Department of
Astronomy, University of Washington, Box 351580, Seattle, WA 98195; 
anderson@astro.washington.edu}

\altaffiltext{3}{Department of Astronomy, Ohio State University,
140 West 18th Avenue, Columbus, OH 43210-1173}

\altaffiltext{4}{Princeton University Observatory, Princeton,
        NJ 08544}

\altaffiltext{5}{Department of Physics and Astronomy, 
University of Pennsylvania, 209 South 33rd Street, Philadelphia, PA 19104}

\altaffiltext{6}{Department of Astronomy, University of California, Los 
Angeles, Los Angeles, CA 90095}

\altaffiltext{7}{Department of Physics and Astronomy, Rutgers, the State 
University of New Jersey, 136 Frelinghuysen Road, Piscataway, NJ 08854}

\altaffiltext{8}{Department of Astronomy, MSC 4500, New Mexico State 
University, P.O. Box 30001, Las Cruces, NM 88003}

\altaffiltext{9}{Kavli Institute for Cosmological Physics, The
University of Chicago, 5640 South Ellis Avenue, Chicago, IL 60637}

\altaffiltext{10}{Center for Particle Astrophysics, Fermilab, P. O. Box 
500, Batavia, IL 60510}

\altaffiltext{11}{Physics Department, University of Notre Dame, 225 
Nieuwland Science, Notre Dame, IN 46556-5670}

\altaffiltext{12}{Apache Point Observatory, P.O. Box 59, Sunspot, NM 
88349}

\altaffiltext{13}{Department of Physics,
        Rochester Institute of Technology, Rochester, NY 14623-5603}

\altaffiltext{14}{Pennsylvania State University, Department of 
        Physics \& Astronomy, 525 Davey Lab, University Park, PA 16802}

\altaffiltext{15}{Department of Astronomy and Astrophysics, The
University of Chicago, 5640 South Ellis Avenue, Chicago, IL 60637}

\altaffiltext{16}{Enrico Fermi Institute,
University of Chicago, 5640 South Ellis Avenue, Chicago, IL 60637}





\begin{abstract}

AM~CVn systems are a select group of ultracompact binaries with the 
shortest orbital periods of any known binary subclass;
mass-transfer is likely from a low-mass (partially-)degenerate secondary 
onto a 
white dwarf primary, driven by gravitational radiation. In the past few 
years, the Sloan Digital Sky Survey (SDSS) has provided five new AM~CVns. 
Here we report on two further candidates selected from more recent SDSS 
data. SDSS J1208+3550 is similar to the earlier SDSS 
discoveries, recognized as an AM~CVn via its distinctive spectrum which is 
dominated by helium emission. From the expanded SDSS Data Release~6 
(DR6) spectroscopic area, we provide an updated surface density estimate 
for such AM~CVns of order $10^{-3.1}$ to $10^{-2.5}$ deg$^{-2}$ for 
$15<g<20.5$. In addition, we present another new candidate 
AM~CVn, SDSS 
J2047+0008, that was discovered in the course of followup of SDSS-II 
supernova candidates. It shows nova-like outbursts in multi-epoch imaging 
data; in contrast to the other SDSS AM~CVn discoveries, its 
(outburst) spectrum is dominated by helium absorption lines, 
reminiscent of KL~Dra 
and 2003aw. The variability selection of SDSS J2047+0008 from the 
300~deg$^2$ of SDSS Stripe 82 presages further AM~CVn discoveries in 
future deep, multicolor, and time-domain surveys such as LSST. The new 
additions bring the total SDSS yield to seven AM~CVns thus far, 
a substantial contribution to this rare subclass, versus the dozen 
previously known.

\end{abstract}





\section{Introduction}

AM Canum Venaticorum (AM~CVn) systems are a select group of ultracompact 
binary systems, with orbital periods extending down to tens of minutes 
\citep[e.g., see reviews by][]{war95,cro04,nel05,ram07}. The ultrashort 
orbital periods (the shortest of any known binary class) suggest that both 
binary components are at least partially degenerate, with a popular model 
invoking accretion driven by gravitational radiation from a fully- or 
semi-degenerate helium-rich donor (perhaps of very low mass) onto a more 
typical white dwarf primary. A hallmark spectroscopic characteristic of an 
AM~CVn is a spectrum dominated by helium lines, and essentially devoid of 
hydrogen.

The unusual character of the prototype, AM~CVn, was discovered 40 years 
ago \citep{sma67,pac67}, but the remarkable class has grown only slowly 
in membership over the intervening decades, e.g., with about a dozen members 
discussed in the review by \citet{nel05}. New cases of AM~CVn 
systems are eagerly sought, as these systems may provide insights into a 
prior common-envelope phase, are discussed as possible SN~Ia progenitors, 
and are plausible sources of gravitational waves 
\citep[e.g., see][]{lie97,liv03,nel04}.

In the last few years the Sloan Digital Sky Survey (SDSS) has discovered 
many new cataclysmic variables \citep{szk02, szk07}, and even five new 
AM~CVn candidates. The first such SDSS AM~CVn system, 
SDSS J1240-0159, was discovered in an early SDSS data release by 
\citet{roe05}. In \citet{and05}, hereafter Paper I, we reported the 
discovery of four additional AM~CVn candidates (including the first 
eclipsing AM~CVn) from the then-available SDSS spectroscopic database. All 
five of those earlier SDSS AM~CVn candidates showed optical spectra 
dominated by strong helium emission, and this characteristic was essential 
in their recognition in the SDSS spectroscopic database.

In this paper, we report on another similar case, SDSS 
J120841.96+355025.2 (hereafter, SDSS J1208+3550), also recognized from 
SDSS spectra because of its strong helium emission. We also provide an 
updated surface density estimate for such emission-line AM~CVns based on 
our expanded DR6 spectroscopic search area.
Additionally, we report a second new AM~CVn candidate, SDSS 
J204739.40+000840.3 (hereafter, SDSS J2047+0008), that was discovered via 
time-domain imaging data as part of the SDSS-II supernova 
survey \citep{fri07}. As initially announced in a telegram by 
\citet{pri06}, SDSS J2047+0008 shows novae-like outbursts in multi-epoch 
SDSS imaging. Moreover, the optical spectrum of SDSS J2047+0008 at 
outburst shows helium {\it absorption} lines, in contrast to the spectra 
of the other SDSS discoveries, and more similar to some other well-known 
AM~CVn systems such as KL~Dra and 2003aw \citep{jha98,fil03}. 
Including the two additions discussed herein, the total 
SDSS yield thus far is seven new AM~CVn candidates, a substantial 
contribution to expanding the membership of this rare subclass, compared 
to the approximately dozen cases known prior to SDSS.

\section{SDSS J120841.96+355025.2, and a Search of SDSS DR6 for Similar
AM~CVn Candidates}

The SDSS is a multi-institutional and international project creating an 
optical digital 
imaging and spectroscopic data bank of a region approaching 
$\sim10^4$~deg$^2$ of sky, primarily centered on the north Galactic polar 
cap. Imaging and spectroscopic data are obtained by a special purpose 2.5m 
telescope, at Apache Point Observatory, New Mexico, equipped with a 
large-format mosaic camera that can image $\sim10^2$~deg$^2$ per night in 
5 filters ($u,g,r,i,z$), along with a multifiber spectrograph that obtains 
640 spectra within a 7~deg$^2$ field.  The imaging database is used to 
select objects for the SDSS spectroscopic survey, which includes 
($\lambda/\Delta\lambda\sim1800$) spectrophotometry covering 3800-9200\AA\ 
for $10^6$ galaxies, $10^5$ quasars, and $10^5$ stars. Technical details 
on SDSS hardware, software, and astrometric, photometric, and spectral 
data may be found in a variety of papers, e.g., \citet{fuk96}, 
\citet{gun98}, \citet{lup99}, \citet{yor00}, \citet{hog01}, \citet{sto02}, 
\citet{smi02}, \citet{pie03}, \citet{ive04}, and \citet{gun06}. A 
description 
of the most recent SDSS Public data release (Data Release 6; hereafter, 
DR6) is given by \citet{ade08}.

As discussed in Paper I, the SDSS collaboration 
includes a ``Serendipity Working Group'' which is engaged in selection 
and examination of a large number of atypical SDSS spectra; such 
spectroscopic  
selection and visual examinations have led to the initial discovery of the 
bulk of the previous SDSS AM~CVns. Their distinctive spectra, similar to 
cataclysmic variables but with helium rather than hydrogen emission lines, 
drew our special attention to these cases.

Similarly, SDSS J1208+3550, the 
first new AM~CVn candidate discussed in this paper, was initially found as 
a serendipitous discovery during a search of the DR5 SDSS spectroscopic 
database for DQ white dwarfs \citep{hal05}.\footnote{This object was 
subsequently independently recovered by others both inside and evidently 
outside the SDSS collaboration; for example, there is a point labeled 
`J1208' in Figures 2 and 4 of \citet{roe07}.} The discovery spectrum of 
SDSS J1208+3550 is 
displayed in Figure 1, and strongly resembles the previous SDSS AM~CVn 
finds, with prominent broad emission at He~I 3888, 4026, 4471, 
4921, 5015, 5875, 6678, 7065, and 7281\AA; for example, the 
equivalent width and full width at half maximum of He~I $\lambda$5875 are, 
respectively, 28 \AA\ and 1600 km~s$^{-1}$. There is also weaker HeII 
$\lambda$4686 emission. Though the signal to noise is modest, the Figure 1 
spectrum again suggests multi-peaked helium emission, presumably from an 
accretion disk. As with most previous SDSS AM~CVn discoveries, SDSS 
J1208+3550 was targeted for a spectrum because of its unusually blue color 
in SDSS imaging data, having been selected by several independent (but 
overlapping) target selection methods including ``hot-standard star", 
``quasar", ``serendipity", and ``white-dwarf" algorithms. Basic 
astrometric and photometric data from SDSS are provided in Table~1, though 
of course AM~CVns may be variable, and so the tabulated SDSS 
J1208+3550 magnitudes are 
only 
indicative of its character at the epoch of the SDSS observations.

Because visual and other serendipitous searches of spectra, 
such as those which initially revealed SDSS J1208+3550, 
could be biased and incomplete, we 
have also algorithmically sifted through the SDSS DR6 
spectroscopic database for any further new SDSS AM~CVn candidates showing 
strong helium emission. This algorithmic search is directly analogous to 
that discussed in detail in Paper I (which was based largely on the 
smaller DR4 spectroscopic area). In the current paper, this algorithmic 
search was applied to DR6 plates encompassing more than  
7000 
square degrees and a million spectra.

In particular, we queried the DR6 spectroscopic database to return a list 
of objects with spectra for which the SDSS pipeline data reduction 
algorithms \citep{sto02} found any emission line with equivalent width 
$>3$\AA, with wavelength centered within 20~\AA\ windows around either 
HeII $\lambda$4686 or HeI $\lambda$5875. These are typically the highest 
equivalent-width lines among the Paper I AM~CVn candidates (initially 
found in the visual search of earlier SDSS spectroscopic plates). The 
choice of minimum equivalent width in this algorithmic search, of course, 
limits such discoveries to AM~CVns with helium emission lines, but does 
encompass all previous SDSS AM~CVn candidates: all earlier SDSS cases 
found in visual searches of spectra have EW$>13$\AA\ for one of the above 
two helium emission lines. The algorithmic SDSS spectral database queries 
initially returned a list
of 23,100 spectra potentially having such emission at either 5875 
\AA\ or 4686 \AA. Each algorithmically-chosen spectrum was then reviewed 
by eye to assess whether it might be an AM~CVn candidate.
This algorithmic search process recovered SDSS J1208+3550 (originally 
in DR5) and all five previously published 
SDSS AM~CVn candidates \citep{roe05,and05}; however, no additional 
convincing AM~CVn candidates were
identified in DR6 based on this algorithmic emission-line search.

That the known AM~CVns occupy small and relatively 
sparsely-populated regions of SDSS color-color diagrams (Figure 2) also 
permits us a secondary search technique for AM~CVn candidates that 
might have slipped through the pipeline emission-line algorithmic 
search discussed above. A total of thirteen AM~CVns fall 
within the SDSS 
DR6 imaging survey region, among which are ten whose magnitudes lie in the 
range $15<g<20.5$; this is an approximate range in which SDSS provides 
high quality imaging photometry and spectroscopy. We construct a 
3-dimensional (hereafter, 3D) box in $u-g$, $g-r$, and $r-i$ multicolor 
space that encompasses those ten AM~CVns; we enlarge the boundaries of 
this box a little in each dimension, as shown in Figure 2, to encompass 
not just the SDSS colors of these ten AM~CVns, but their colors plus 
$3\sigma$ color errors as well. (In constructing this box, we do not 
consider three additional AM~CVns in the SDSS area that were either too 
bright or too faint for reliable photometry. Nor do we consider the $i-z$ 
color distribution of the AM~CVns, as the $z$-band photometric errors are 
large for most of these blue AM~CVns imaged in SDSS).
We then visually examined the 8300 objects with spectra in SDSS 
DR6, whose colors in SDSS imaging fall within this 3D multicolor box. This 
additional perusal of SDSS DR6 color-selected spectra again recovered 
all five previously published SDSS AM~CVns, but revealed no 
additional strong emission-line AM~CVn candidates aside from SDSS 
J1208+3550. Thus, SDSS
J1208+3550 (originally from DR5) is the solitary new emission-line
candidate AM~CVn discussed herein.

Following our initial 2005 discovery of SDSS J1208+3550, in 2006-2007 we 
then obtained follow-up exploratory optical lightcurve and X-ray flux 
observations of this new AM~CVn system. We used the SPIcam CCD imager with 
an SDSS $g$-band filter on the ARC 3.5m to obtain optical lightcurve 
information from brief time-series sequences (spanning about 1-2 hours 
each) on four distinct nights: 2006 Feb. 1, 2006 Mar. 3, 2006 
Apr. 1, and 2007 Jan. 12 (all UT). Several of these nights were not 
photometric, but differential photometry was obtained relative to (the 
same) three brighter comparison stars in each image. The time resolution 
between individual exposures was in the range 1.4--2.4~minutes for all 
nights. Unfortunately, these data reveal no convincing periodic optical 
modulation for periods in the range of about 4-40 minutes, and with an 
amplitude limit of a few hundredths of a magnitude. Longer timespan 
observations would be especially useful to obtain similar sensitivity to 
photometric modulations at (plausibly) longer orbital periods.

We also obtained a $\sim$3000s X-ray exposure of SDSS J1208+3550 from 
the {\it Chandra} X-ray Observatory on 2007 Mar. 23 (UT). These data 
were collected in VFAINT mode allowing us to utilize the full 5$\times$5 
pixel event island and thus reduce ACIS particle background. We 
reprocessed the Level~1 events file in order to leverage this additional 
information and thereby filter for hot pixels, cosmic rays, 
background events, bad grades, and good time intervals. SDSS J1208+3550 
is strongly detected in our {\it Chandra} data, with 45 counts in a medium 
energy ($0.5-4.5$~keV) X-ray band; background contributions are evaluated 
with a monoenergetic ($0.5$~kev) exposure map. For an assumed blackbody 
X-ray spectrum with $kT = 0.7$~keV, e.g. similar to that inferred for 
ES~Cet \citep{str04a}, these {\it Chandra} medium band data imply an 
unabsorbed $0.2-10$~keV flux of about $1.0\times10^{-13}$ 
erg~s$^{-1}$~cm$^{-2}$. This measured X-ray flux for SDSS J1208+3550 is 
reassuringly similar (within about a factor of 2) to that we predicted 
based on the typical X-ray to optical flux ratios of other AM CVns.

\section{SDSS J204739.40+000840.3, a Time-Domain Selected AM~CVn 
Candidate} 

Since the completion of Paper I, an entirely new 
opportunity for 
AM~CVn searches has emerged from SDSS, and here we also report
a new AM~CVn, SDSS J2047+0008,  
discovered in that very different manner. SDSS J2047+0008 
(originally known internally as ``candidate 15204")
was instead
discovered in the 
course of the SDSS-II 
supernova search
(Frieman et al. 
2007), which surveys the 300 deg$^2$ SDSS ``Stripe 82" 
region via multi-epoch imaging.
Basic data on SDSS J2047+0008 are displayed in Table 1, which reflect the 
object's state (in outburst) on UT 2006 Oct. 12 (MJD 54020.125); its 
color on 
this epoch also falls nicely within the 3D multicolor box found for other 
AM~CVns shown in Figure 2. 

This new object is markedly variable. As may be discerned from the Figure 
3 long-term lightcurve, 
SDSS J2047+0008 has been imaged with SDSS-I and SDSS-II Stripe 
82 observations at
60 epochs over an 8-year timespan \citep{ive07,hol08}, but was unnoticed 
until
its outburst in Fall of 2006, when caught as part of the SDSS-II supernova 
discovery program. 
The large amplitude variations 
strongly suggests 
nova-like outbursts in this system.

The SDSS-II lightcurve of SDSS J2047+0008 shown in Figure 4
has been expanded to 
highlight the 
Fall 2006 discovery and outburst observations. SDSS J2047+0008
(not too 
surprisingly) appears to have a stronger ultraviolet excess in its 
brighter state. A potentially interesting aspect of the lightcurve is 
that this object may show some cycling between high and intermediate 
states while fading; such behavior, for example, has 
been reported for the (outbursting) AM~CVn system V803 Cen 
\citep{pat00,kat04}.

Prompted by its dramatic variability in SDSS-II imaging, an optical 
spectrum of SDSS J2047+0008 was taken on UT 2006 Oct. 19 (MJD 54027) 
with the MDM 2.4-m Hiltner telescope (plus OSU Boller and Chivens 
Spectrograph), and is shown in Figure 5. There is clearly a blue 
continuum with shallow helium absorption lines of He I 3820, 3867, 
3927, 4009, 4026, 4144, 4388, 4471, 4921, and perhaps 5875 \AA. This 
spectrum is 
unlike traditional novae, but similar to the other objects such as KL Dra 
(also known as 1998di; Jha et al. 1998) and 2003aw (Filippenko \& Chornock 
2003), that were also noticed as hydrogen 
deficient novae, and subsequently classified as likely AM~CVns.
 
The absorption-line spectral character of this new 
variability-selected AM~CVn system, SDSS J2047+0008, is in strong contrast 
to the 
other SDSS discovered AM~CVns, which were all recognized
in the SDSS database by virtue of their helium-dominated
emission line spectra. 
The SDSS databases (including the SDSS-II 
Stripe 82
data) may include interesting additional AM~CVns 
similar to SDSS J2047+0008,
still awaiting recognition.

A followup lightcurve of SDSS J2047+0008 in outburst, and with time 
resolution potentially sensitive to a strong orbital modulation, was 
obtained with the WIYN telescope and MiniMosaic CCD array on 2006 Nov. 1 
(UT). Exposures were 300 seconds long through a Harris $R$-band filter, 
and 
each exposure had a two minute readout time, yielding 19 images over a 2.5 
hour time-span. Differential photometry was performed by comparison to 
nearby stars with known SDSS calibrations. A star 15$''$ to the 
southeast of SDSS J2047+0008 is estimated to have a $R$-band magnitude of 
16.46$\pm 0.02$ mag by converting from SDSS magnitudes. From this 
approximate calibration, the average magnitude of SDSS J2047+0008 over the 
WIYN observing run was found to be $R=20.58$, but with significant 
variations of 0.2 magnitudes over the 2.5 hour span. Additional lightcurve 
data 
taken at good time-resolution may be useful in determining an orbital 
period; however, a quality lightcurve outside of outburst may prove 
difficult, given the faintness of the object in quiescence (Figure~3).

\section{Discussion}

Our spectroscopic search for emission-line AM~CVn candidates
from SDSS now includes spectroscopic plates extending through DR6,
and provides some updated, though approximate, constraints on the 
surface density of AM~CVns. These 
constraints remain uncertain as SDSS spectroscopy (purposefully  
emphasizing the main galaxy and quasar surveys) is highly incomplete 
and non-uniform across the sky even for emission-line AM~CVn candidates
(see details in Paper I).

The main SDSS spectroscopic survey, along with the SDSS-II SEGUE survey, 
include spectroscopic plates which encompass 7400 square degrees of sky 
and 1.3 million spectra. SDSS has discovered six emission-line AM~CVn 
candidates in this region, including SDSS J1208+3550 presented herein; all 
six were initially selected as interesting 
for spectroscopic follow-up based on their blue colors in SDSS imaging, 
with subsequent recognition as AM~CVns in the SDSS spectroscopic database 
because of their helium-dominated emission line spectra (Roelofs et al. 
2005; Paper I; this paper). All six of these had $15<g<20.5$ at the epochs 
of both SDSS imaging and SDSS spectrophotometry, and this is also a 
plausible range for both high-quality SDSS imaging and spectra. (For 
example, SDSS spectroscopy is restricted to $m>15$ to avoid fiber 
cross-talk; and for blue stellar objects, $m<20.5$ is the typical 
faintest limit for routine spectroscopic target selection.) Thus, an 
approximate, conservative lower limit on the surface density of AM~CVns is 
$> 1/1200$~deg$^{-2}$ for $15<g<20.5$.

To proceed further, albeit less securely, requires an estimate of 
the incompleteness of SDSS spectroscopy for similar 
emission-line AM~CVn 
candidates.
The incompleteness even for very similar AM~CVn candidates arises from
several complicating factors, including: 
the emission-line AM~CVn candidates found 
in SDSS were selected for spectroscopy by several distinct target
selection algorithms, each differing in their specific color-selection 
criteria  
and limiting magnitudes; and, some of these 
target selection algorithms actually receive spare spectroscopic fibers
only rarely and non-uniformly over the sky, 
when the main
galaxy and quasar surveys do not consume all available fibers for a given
spectroscopic plate. 

But, as a rough estimate of such spectroscopic incompleteness for similar 
AM~CVn candidates, we consider again the same 3D box in $u-g$, $g-r$, 
$r-i$ 
multi-color imaging space described in section 2 (also see 
Figure 2). In the DR6 imaging area
there are about 2.97~deg$^{-2}$ stellar objects with such colors and 
$15<g<20.5$; in the DR6 spectroscopy area, the surface density of such 
objects actually having SDSS spectra taken is about 1.14~deg$^{-2}$.
Thus, SDSS spectroscopy would be of order 38\% complete for such 
objects, naively assuming uniform spectral coverage throughout the 3D 
box.

Alternately, one might consider a somewhat smaller box in multicolor space 
that extends only to objects as red as about 
$u-g<0.11$; as may be discerned from Figure 2, this smaller 3D box 
excludes only one of the (ten) known AM~CVns with reliable photometry in 
DR6. In the DR6 imaging area, such a smaller 3D multicolor box includes 
a typical surface density of about 1.04~deg$^{-2}$ of such objects. The 
surface density of such SDSS objects also having spectroscopy is about 
0.40~deg$^{-2}$, suggesting (naively) a similar SDSS spectroscopic overall 
completeness of order 39\%. 
In fact, the situation is more complicated, as such spectroscopic 
completeness depends on various magnitude limits and may vary across even 
this smaller multicolor box, due to the differing target selection 
algorithms used. More elegant estimates may be tied to AM~CVn binary 
period and temperature/color relations; see \citet{roe07} for an excellent 
example of this approach.

As an intermediate improvement on our simple estimate above, we have 
considered the 
SDSS spectroscopic 
completeness (accounting for differences in DR6 imaging and 
spectroscopic survey areas) in these 
same 3D multicolor boxes for each 0.5 magnitude bin in the range 
$15.0<g<20.5$.
Our spectroscopic completeness estimates are very similar when
considering either the larger or smaller 3D multicolor boxes, so the 
precise box color boundaries may not be critical; completeness 
(for morphologically stellar objects) ranges from a low of 10-20\% at the 
very 
brightest and  faintest ends, maximizing at just under 
70\%  spectral completeness at $g=18-19$. Within the faintest four bins 
spanning 
$g=18.5-20.5$, 
where the great bulk of SDSS 3D color-selected candidates would in 
any case fall, and 
which also encompasses 
all the emission-line selected AM~CVns from SDSS, the lowest 
completeness estimate is 24\%. We thus adopt the latter 
value as a 
representative lower 
limit to 
the SDSS spectroscopic completeness within the 3D boxes shown in Figure 2. 
This updated completeness estimate is also quite similar to the $\sim$20\% 
we suggested in 
Paper~I, and even to the more sophisticated completeness estimates 
of 
\citet{roe07} in a similar magnitude regime (and for a wide range in 
AM~CVn binary periods in their model).

Broadly then, the surface density of AM~CVns with 
$15 \lesssim g \lesssim 20.5$, similar 
to those with strong helium emission as found in the SDSS 
spectroscopic database, might be expected
to be approximately in the range from $10^{-3.1}$ deg$^{-2}$ (with no 
spectroscopic completeness 
correction) to about  
$10^{-2.5}$ deg$^{-2}$ (adopting representative spectroscopic 
completeness of
$>24\%$). Of course there are 
additional AM~CVns 
aside 
from those typically showing strong emission lines; indeed, 
SDSS J2047+0008 discussed in section 3 is
the most recent example.

In one commonly discussed scenario (e.g., Nelemans 2005), the youngest 
members of the 
AM~CVn class are thought to be those with the highest mass-transfer
rates and shortest periods ($<$20 minutes), and these ``high-state" 
systems show predominantly 
helium in absorption in their optical spectra. In that scenario, they 
then evolve
into somewhat longer periods as the binary widens and the mass 
transfer rate continues to decrease; in this intermediate stage, 
perhaps with 20-40 
minute periods typical, the objects may show nova-like outbursts, and 
display both
high-state (absorption) and low-state (emission) spectra. Then finally 
there is a late stage (perhaps with typical 
40-60 minute binary periods) in which AM~CVns are primarily low-state 
systems
with predominantly emission-line spectra. Other scenarios similarly place
an approximate dividing line between mainly emission-line and 
mainly absorption-line 
systems at binary periods of about 30~minutes \citep{roe07}.
The relative populations in 
these various stages are not yet well constrained empirically,
but at least some models \citep{roe07} predict significantly 
fewer of
the shorter-period systems, and so the surface 
density of all AM~CVns might be comparable to the estimate above 
based on 
emission-line AM~CVns only.

At first glance these fiducial surface density limits are also compatible 
with 
our discovery of the one variability-selected case, SDSS J2047+0008, from 
the SDSS Stripe 82 survey region, given: (i) the uncertainty in the above 
surface density estimates; (ii) the 300 deg$^2$ areal coverage of 
time-domain imaging of the Stripe 82 supernova survey region; (iii) the 
faintness in quiescence (often $g>22$) of SDSS J2047+0008, a regime for 
which presumably the surface densities are even higher; and, (iv) the 
large uncertainty in the relative populations among high, intermediate, 
and low-state AM~CVns.

Despite recent down-sizing in population model estimates for AM~CVns based 
on 
essentially this same set of SDSS objects \citep{roe07}, our
current surface density estimates suggest that upcoming large-area 
multicolor plus time-domain surveys 
such as LSST---with their excellent time-sampling and color-range, and 
extending to 
fainter 
magnitudes than SDSS---may still provide an excellent opportunity to 
discover dozens or even hundreds 
of additional AM~CVns. Such a large area,
multi-color plus time-domain sample might especially extend
uniform surveys to the intermediate and high-state AM~CVn populations that 
(often) lack emission lines, and which may be much more difficult to 
efficiently identify in multicolor-only surveys.

In any case, along with SDSS J1208+3550 and J2047+0008 presented herein, 
SDSS-I and SDSS-II have thus 
far yielded a total of seven new AM~CVn candidates 
(Roelofs et al. 2005; Anderson et al. 2005; this paper). SDSS is
thus providing a substantial expansion to this rare object class, compared 
to the approximately dozen AM~CVns previously known.

\acknowledgments

    Funding for the SDSS and SDSS-II has been provided by the Alfred P. Sloan 
Foundation, the Participating Institutions, the National Science Foundation, the 
U.S. Department of Energy, the National Aeronautics and Space Administration, the 
Japanese Monbukagakusho, the Max Planck Society, and the Higher Education Funding 
Council for England. The SDSS Web Site is http://www.sdss.org/.

    The SDSS is managed by the Astrophysical Research Consortium for the 
Participating Institutions. The Participating Institutions are the American Museum 
of Natural History, Astrophysical Institute Potsdam, University of Basel, 
University of Cambridge, Case Western Reserve University, University of Chicago, 
Drexel University, Fermilab, the Institute for Advanced Study, the Japan 
Participation Group, Johns Hopkins University, the Joint Institute for Nuclear 
Astrophysics, the Kavli Institute for Particle Astrophysics and Cosmology, the 
Korean Scientist Group, the Chinese Academy of Sciences (LAMOST), Los Alamos 
National Laboratory, the Max-Planck-Institute for Astronomy (MPIA), the 
Max-Planck-Institute for Astrophysics (MPA), New Mexico State University, Ohio 
State University, University of Pittsburgh, University of Portsmouth, Princeton 
University, the United States Naval Observatory, and the University of Washington.

	DH acknowledges support from the NASA Harriett G.\ Jenkins 
Predoctoral Fellowship. Support for this work was provided by the National 
Aeronautics and Space Administration through Chandra Award Number 
GO7-8028X issued by the Chandra X-ray Observatory Center, which is 
operated by the Smithsonian Astrophysical Observatory for and on behalf of 
the National Aeronautics Space Administration under contract NAS8-03060.

\clearpage



\begin{figure}
\plotone{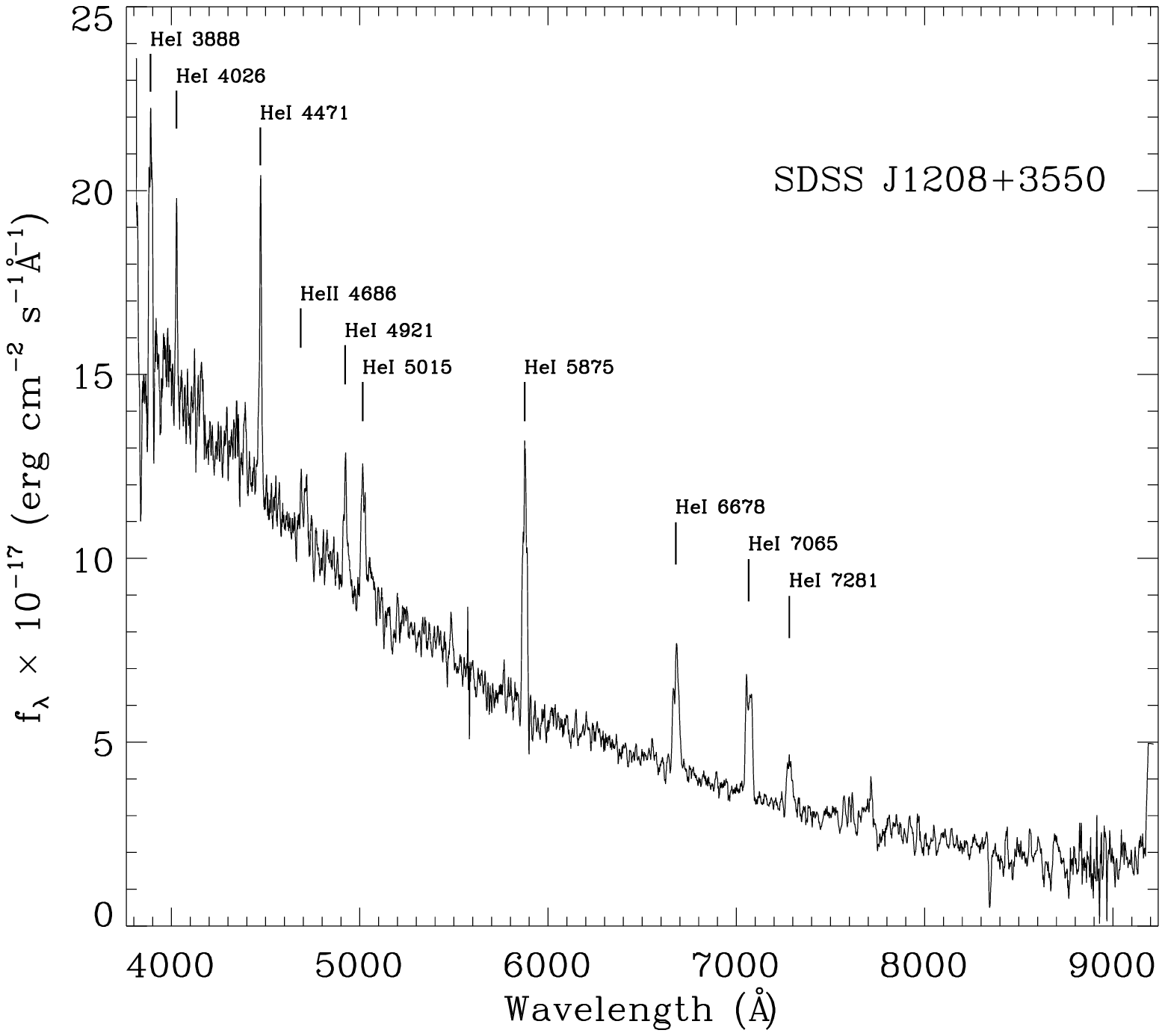}
\vspace*{1cm}
\caption{The SDSS discovery 
spectrum of SDSS J1208+3550,
displayed here with a 7-point boxcar smooth. This AM~CVn
candidate was initially recognized in the SDSS DR5 spectroscopic database 
due to its 
helium-dominated emission line spectrum, which is very similar 
to the earlier SDSS AM~CVn discoveries. 
\label{fig1}}
\end{figure}

\begin{figure}
\voffset=-0.5in
\epsscale{1.1}
\centerline{\plotone{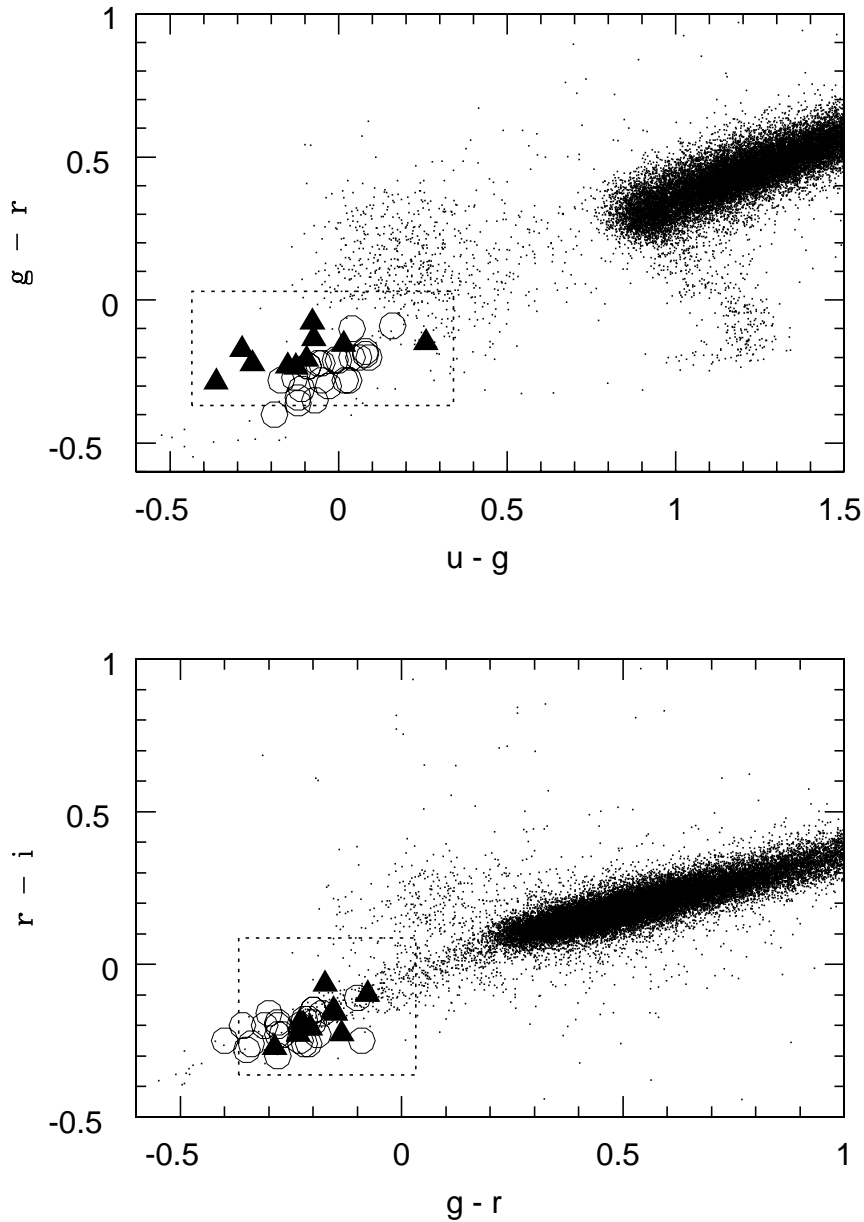}}
\voffset=-0.5in
\caption{SDSS color-color diagrams highlighting AM~CVn
systems. The small black points show the 
colors of random stellar objects having 
reliable photometry, derived from 100~deg$^2$ of SDSS. 
Over-plotted are the SDSS colors
of ten AM~CVns (solid, black triangles) with reliable photometry, that fall within
SDSS DR6 imaging area.
For additional comparison, the colors of SDSS DB white dwarfs from 
\citet{har03} are also plotted (open circles).
The AM~CVn systems occupy a relatively
sparse and distinct portion of SDSS multicolor space,
as depicted by the dotted boxes.
\label{fig2}}
\end{figure}

\begin{figure}
\plotone{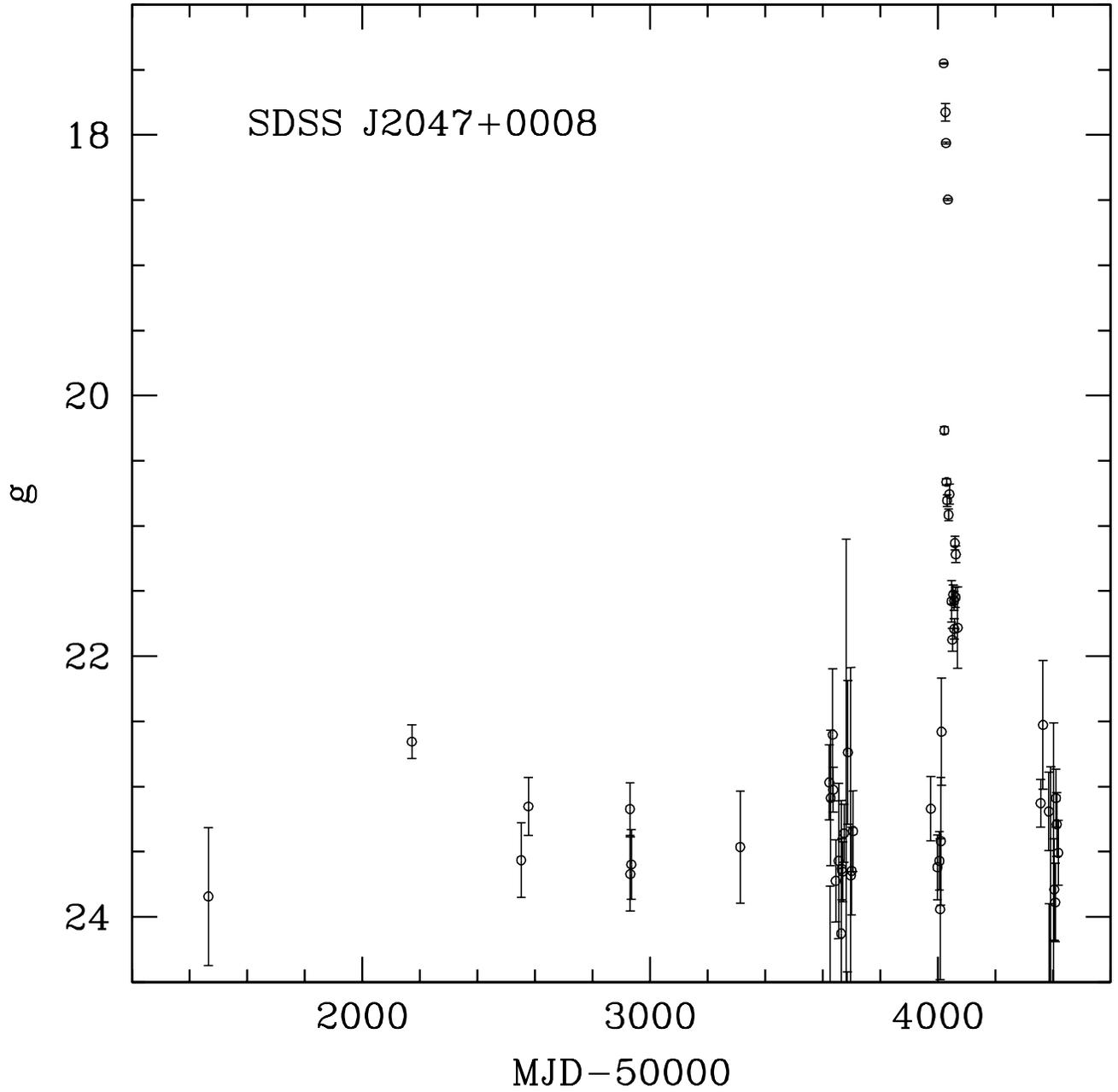}
\caption{The SDSS Stripe 82 long-term $g-$band (asinh magnitudes 
are plotted) lightcurve 
of SDSS J2047+0008. 
This object first attracted attention because of its marked
outburst (near MJD 54020) in Fall of 2006
in multi-epoch imaging data for
the SDSS-II supernova survey.}
\label{fig3}
\end{figure}

\begin{figure}
\plotone{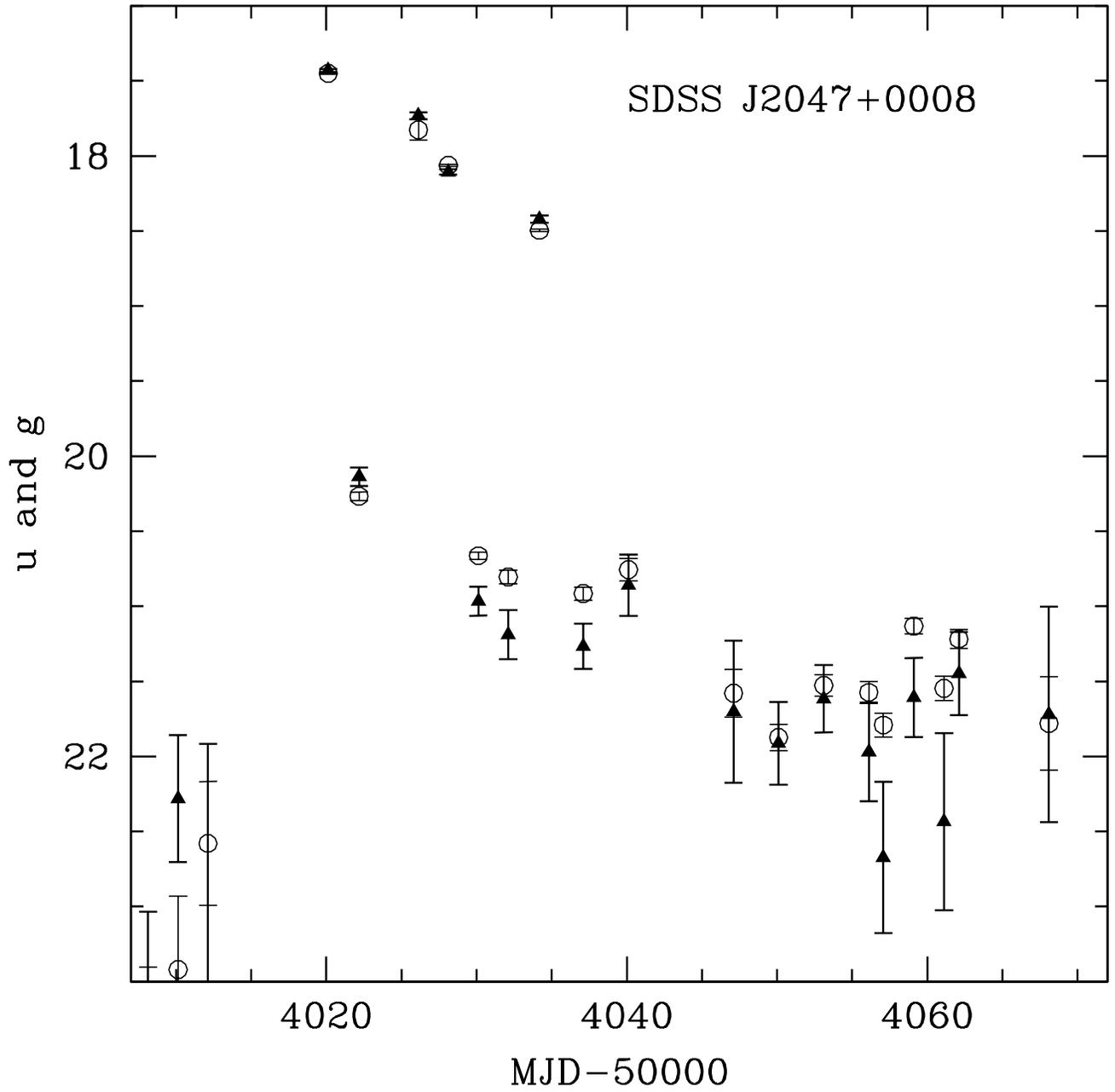}
\caption{A portion of the SDSS Stripe 82 $u$-band (filled triangles) and 
$g-$band (open circles) lightcurves
of SDSS J2047+0008, highlighting the Fall 2006 outburst. (The plotted 
points are again asinh magnitudes.)}
\label{fig4}
\end{figure}

\begin{figure}
\plotone{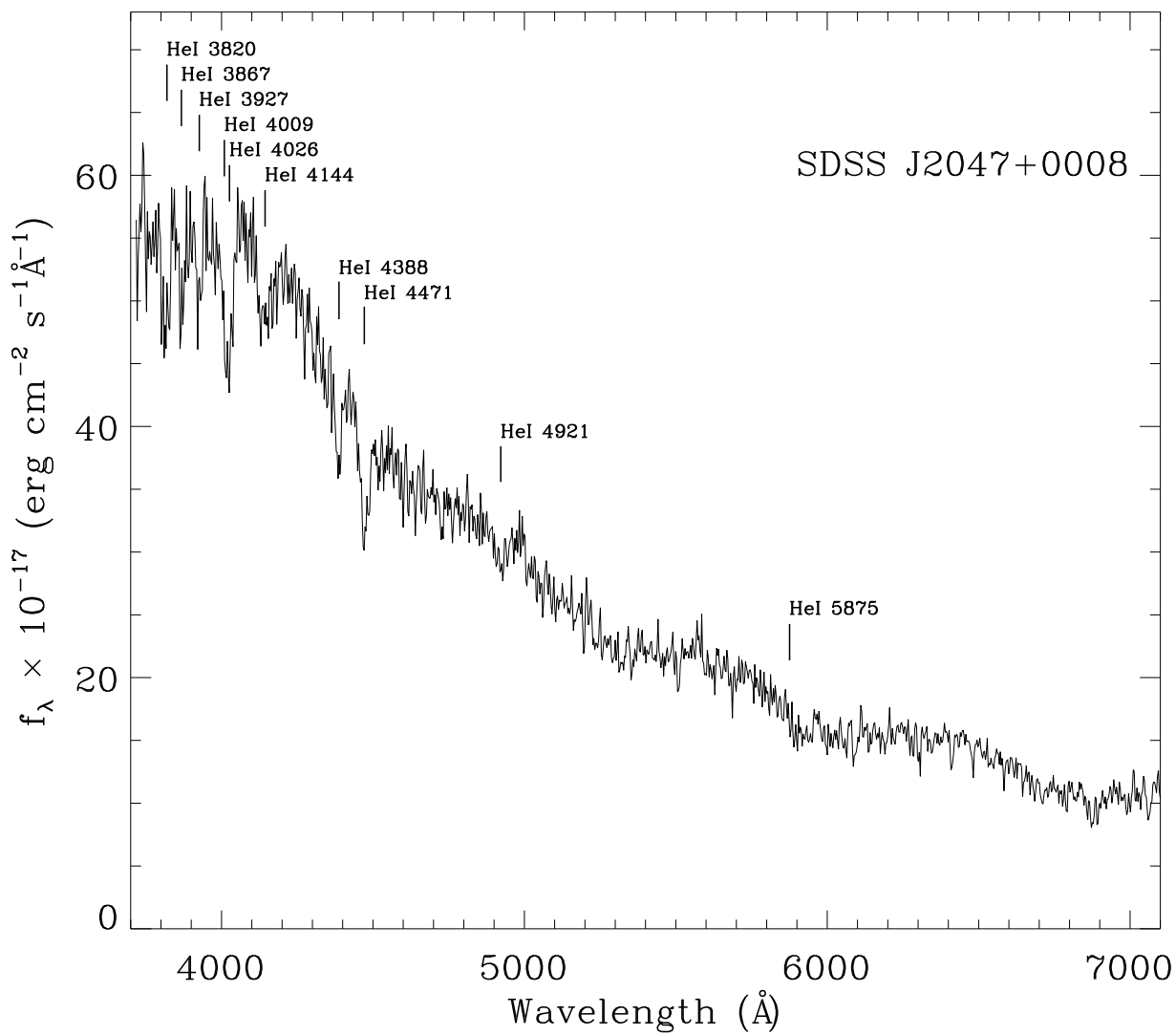}
\caption{The MDM spectrum of variability-selected object SDSS J2047+0008. 
This optical spectrum 
confirms SDSS J2047+0008 as a strong AM~CVn candidate.
The spectrum is dominated by helium in
absorption, and was
taken (on MJD 54027) in a high, outbursting state. The 
absorption dominated
helium spectrum is 
in contrast to the other SDSS-discovered AM~CVns which have
helium in emission, but similar
to some other well-known AM~CVns, such as KL Dra and 2003aw.}
\label{fig5}
\end{figure}






\clearpage

\begin{deluxetable}{lcccccccccc}
\tabletypesize{\footnotesize}
\tablecaption{Some Basic Data on SDSS J1208+3550 and J2047+0008 }
\tablewidth{0pt}
\tablecolumns{11}
                                                                                                                                                       
\tablehead{ \colhead{RA,Dec name} &
   \colhead{{\it u}} & \colhead{{\it g}} &
   \colhead{{\it r}} &
   \colhead{{\it i}} & \colhead{{\it z}} &
   \colhead{Epoch}}
                                                                                                                                                       
\startdata
J120841.96+355025.2 & 18.80 & 18.79 & 18.94 & 19.09 & 19.17 & 2004 Apr. 16 (UT)\\
J204739.40+000840.3 & 17.43 & 17.45 & 17.76 & 18.02 & 18.27 & 2006 Oct. 12 (UT)\\
\enddata
                                                                                                                                                       

\end{deluxetable}

\end{document}